\documentclass[preprint,12pt]{elsarticle}

\usepackage{amsmath,amssymb,bm,mathtools}
\usepackage{booktabs}
\usepackage{microtype}
\usepackage{hyperref}

\journal{Physics Letters B}

\begin{document}

\begin{frontmatter}

\title{Three-Body Holonomy and $SU(3)\to SU(2)\times U(1)$ Symmetry Breaking}

\author[inst1]{Tadashi Yoshikawa\corref{cor1}}
\cortext[cor1]{Corresponding author.}
\affiliation[inst1]{organization={Nagoya Aoi University},
            city={Nagoya},
            country={Japan}}

\begin{abstract}
We show that a three-body holonomy can select the conventional 
$SU(3)\to SU(2)\times U(1)$ breaking direction. 
Three particles are special in the present construction: after removal 
of the center-of-mass coordinate they are the minimal one-dimensional few-body system 
with a two-dimensional relative space, allowing a closed cycle 
around the triple-coincidence point once pairwise contacts 
are encoded by matching conditions.  A cyclic three-state mass operator gives
\[
M^{[k]}=M_0+2\Delta\cos\!\left(\frac{\theta_3+2\pi k}{3}\right),\qquad k=0,1,2.
\]
At $\theta_3=0$ the spectrum splits as $3\to2+1$, and the induced mass operator 
is proportional to the Cartan generator $T_8$.  Its commutant 
is therefore $SU(2)\times U(1)$.  In a trinification-like embedding, 
distinct left- and right-sector holonomies leave the Cartan directions entering
\[
Y=T_{3R}+\frac{1}{\sqrt3}(T_{8L}+T_{8R}),
\]
which reproduces the Standard-Model hypercharges for the usual trinification matter 
assignment.  The construction is a toy mechanism for selecting the breaking direction; 
additional dynamics are required for rank reduction and vacuum selection.
\end{abstract}


\end{frontmatter}

\section{Introduction}
Relativistic few-body Dirac systems in $1+1$ dimensions provide an exactly 
tractable setting for studying nonperturbative many-body 
structures \cite{Munakata1990,Sakamoto1993,SakamotoNakanoYoshikawa1993}.  
A key feature of the present construction is that three 
is the minimal nontrivial particle number.  
For two particles the relative coordinate is one-dimensional, 
so there is no closed winding around the coincidence point.  
For three particles, removing the center-of-mass coordinate leaves 
a two-dimensional relative configuration space.  
Once the pairwise contact interactions are represented by matching conditions 
across the coincidence lines, 
the triple-coincidence point can support a closed sector cycle 
and an associated holonomy.  The three coincidence lines divide 
the relative plane into six ordering sectors, which at fixed parity reduce 
to a cyclic three-state space.

On this reduced space, consider a Hermitian mass operator 
with a gauge-invariant phase $\theta_3$ accumulated 
around the three-step cycle.  Its eigenvalues are
\begin{equation}
M^{[k]}=M_0+2\Delta\cos\!\left(\frac{\theta_3+2\pi k}{3}\right),
\qquad k=0,1,2,
\label{eq:spectrum}
\end{equation}
where $\Delta$ measures cyclic mixing.  
The phase is the finite-dimensional analogue of a holonomy 
around a closed path \cite{AharonovBohm}.

This Letter focuses on one consequence of Eq.~\eqref{eq:spectrum}: at $\theta_3=0$ the triplet splits as $3\to2+1$, and the splitting operator is directed along $T_8$.  Hence the unbroken continuous subgroup is $SU(2)\times U(1)$.  The $SU(3)$ here is the continuous basis symmetry of a degenerate three-dimensional internal space, not the discrete group $C_3$ itself; non-Abelian connections on degenerate subspaces are familiar from the Wilczek--Zee construction \cite{WilczekZee}.  We then show that, in a trinification-like embedding, the surviving Cartan directions contain the Standard-Model hypercharge generator.  Our claim is limited to the selection of the breaking direction, not to a complete derivation of the Standard Model.

\section{Three-body cyclic space and holonomy}
For three coordinates $x_1,x_2,x_3$, introduce Jacobi variables
\begin{equation}
r=\frac{x_1-x_2}{\sqrt{2}},\qquad
s=\frac{x_1+x_2-2x_3}{\sqrt{6}}.
\end{equation}
The three coincidence conditions $x_1=x_2$, $x_2=x_3$, and $x_3=x_1$ become 
three straight lines through the origin in the $(r,s)$ plane.

\subsection{Why three is special}
The number three is not introduced here merely as a phenomenological multiplicity.  
In one spatial dimension an $N$-particle system has an $(N-1)$-dimensional 
relative configuration space after removal of the center of mass.  
For $N=2$ this space is one-dimensional; the coincidence point separates the line, 
and there is no continuous closed path that winds around it.  
The case $N=3$ is therefore the first one in which the relative space is two-dimensional.
The triple coincidence is then a codimension-two point.  
After the pairwise singular interactions are transferred to matching conditions 
on the three coincidence lines, a state can be continued successively 
through the six ordering sectors and return to its starting sector.  
The phase accumulated along this closed sector cycle is the three-body holonomy.

At the same time, the geometry itself produces a second occurrence of three.  
The three coincidence lines generate six ordering sectors, 
and fixing intrinsic parity identifies opposite sectors, giving
\begin{equation}
6\longrightarrow 3.
\end{equation}
Thus three is distinguished twice in the present mechanism: 
it is the minimal particle number that permits the closed holonomy cycle, 
and it is the dimension of the resulting cyclic internal space.  
This statement is specific to the present construction; 
it does not assert that three is uniquely selected in all possible microscopic theories.

These lines divide the relative configuration space into six ordering sectors.  
In a sector of fixed intrinsic parity, opposite ordering sectors are paired, 
leaving three cyclic sectors, which we denote by
\begin{equation}
|0\rangle,\qquad |1\rangle,\qquad |2\rangle.
\end{equation}
The cyclic operator $C_3$ acts as
\begin{equation}
C_3|j\rangle=|j+1\;({\rm mod}\;3)\rangle,
\qquad C_3^3=1.
\end{equation}

A minimal Hermitian effective mass operator carrying a gauge-invariant phase 
around the cycle is
\begin{equation}
\mathcal M_3
=M_0 I_3+
\Delta\left(e^{i\theta_3/3}C_3+e^{-i\theta_3/3}C_3^\dagger\right).
\label{eq:Mcyclic}
\end{equation}
Diagonalizing $C_3$ gives the eigenvalues in Eq.~\eqref{eq:spectrum}.  
The crucial point is that the dependence on $\theta_3$ is collective: 
it is associated with the closed three-step cycle.  
Individual link phases may be redistributed by rephasing the basis states, 
whereas their sum around the cycle, $\theta_3$, is invariant.

\section{$T_8$ selection and $SU(3)\to SU(2)\times U(1)$}
At $\theta_3=0$ one obtains
\begin{equation}
M^{[0]}=M_0+2\Delta,\qquad
M^{[1]}=M^{[2]}=M_0-\Delta.
\label{eq:split}
\end{equation}
After a basis permutation that places the degenerate pair in the first two entries,
\begin{equation}
\mathcal M_3
=M_0 I_3+
\Delta
\begin{pmatrix}
-1&0&0\\
0&-1&0\\
0&0&2
\end{pmatrix}.
\label{eq:diagM}
\end{equation}
Using
\begin{equation}
T_8=\frac{1}{2\sqrt3}
\begin{pmatrix}
1&0&0\\
0&1&0\\
0&0&-2
\end{pmatrix},
\end{equation}
Eq.~\eqref{eq:diagM} becomes
\begin{equation}
\mathcal M_3=M_0I_3-2\sqrt3\,\Delta T_8.
\label{eq:T8mass}
\end{equation}
Thus the holonomy selects the $T_8$ direction in the $SU(3)$ algebra.

Let $T_a=\lambda_a/2$ be the conventional Gell-Mann generators.  
Equation~\eqref{eq:T8mass} obeys
\begin{align}
[\mathcal M_3,T_a]&=0, \qquad a=1,2,3,8,\\
[\mathcal M_3,T_a]&\neq0, \qquad a=4,5,6,7,
\end{align}
for $\Delta\neq0$.  Hence its continuous commutant is
\begin{equation}
S[U(2)\times U(1)]
\simeq \frac{SU(2)\times U(1)}{\mathbb Z_2},
\end{equation}
which we denote schematically by
\begin{equation}
SU(3)\xrightarrow{\;\theta_3=0\;}SU(2)\times U(1).
\label{eq:breaking}
\end{equation}
The result is therefore stronger than a generic three-level splitting: 
the $2+1$ degeneracy picks the same Cartan direction used 
in the conventional $SU(3)\to SU(2)\times U(1)$ embedding.

\section{Trinification-like embedding}
We now consider the trinification product 
symmetry \cite{DeRujula1984,Babu1986,Chalut2002,HetzelStech2015}
\begin{equation}
G_3=SU(3)_C\times SU(3)_L\times SU(3)_R.
\label{eq:tri}
\end{equation}
The present proposal does not require the three factors 
to have the same microscopic origin.  In particular, 
the color-like triplet may originate from a coincidence 
or geometric three-state structure, while the left and right triplets 
may originate from a cyclic holonomy acting on chiral sectors.  
Related $SU(3)^3$ constructions have also been studied with 
a discrete $S_3$ permutation symmetry, including in the context 
of CP and strong-CP physics \cite{Chalut2002}.  What is required 
for the following argument is only that the three factors 
are mutually commuting internal symmetries at the scale 
where Eq.~\eqref{eq:tri} is meaningful.

For the left sector, choose the special holonomy of Eq.~\eqref{eq:split}.  Then
\begin{equation}
SU(3)_L\to SU(2)_L\times U(1)_{8L}.
\label{eq:leftbreak}
\end{equation}
For the right sector, a generic Cartan-valued holonomy can be written as
\begin{equation}
\Omega_R=
\exp\left[i\left(\phi_{3R}T_{3R}+\phi_{8R}T_{8R}\right)\right].
\end{equation}
For generic eigenvalues this leaves
\begin{equation}
SU(3)_R\to U(1)_{3R}\times U(1)_{8R}.
\label{eq:rightbreak}
\end{equation}
Combining Eqs.~\eqref{eq:leftbreak} and \eqref{eq:rightbreak}, 
the surviving Abelian generators are $T_{8L}$, $T_{3R}$, and $T_{8R}$.

A standard hypercharge direction in trinification is \cite{Babu1986,HetzelStech2015}
\begin{equation}
Y=T_{3R}+\frac{1}{\sqrt3}\left(T_{8L}+T_{8R}\right).
\label{eq:Y}
\end{equation}
The electric charge is then
\begin{equation}
Q=T_{3L}+Y.
\label{eq:Q}
\end{equation}
Two orthogonal Abelian directions may for example be chosen as
\begin{align}
X_1&=T_{8L}-T_{8R},\\
X_2&=T_{8L}+T_{8R}-\frac{2}{\sqrt3}T_{3R}.
\end{align}
A further rank-reducing condensate or boundary condition 
must remove $X_1$ and $X_2$ while preserving $Y$.  
A purely Cartan holonomy does not by itself reduce the total rank 
from four to two.  Conventional trinification models accomplish 
the remaining reduction through scalar vacuum expectation values 
or related symmetry-breaking ingredients \cite{Babu1986,CamargoMolina2016}.

\section{Standard-Model hypercharges}
To verify that Eq.~\eqref{eq:Y} selects the desired Abelian direction, 
consider the usual trinification matter assignment \cite{Babu1986,HetzelStech2015}, 
written in left-handed Weyl notation,
\begin{equation}
Q_L\sim(3,3,1),\qquad
Q_R^c\sim(\bar3,1,\bar3),\qquad
L\sim(1,\bar3,3).
\label{eq:matter}
\end{equation}
For the fundamental representation,
\begin{equation}
T_3={\rm diag}\left(\frac12,-\frac12,0\right),\qquad
\frac{T_8}{\sqrt3}={\rm diag}\left(\frac16,\frac16,-\frac13\right),
\end{equation}
and the generators in the anti-fundamental have opposite eigenvalues.

The Standard-Model components then acquire the hypercharges shown in Table~\ref{tab:Y}.

\begin{table}[ht]
\centering
\caption{Hypercharges obtained from Eq.~\eqref{eq:Y}. 
Right-handed Standard-Model fields are displayed with their conventional charges 
rather than as left-handed charge-conjugate fields.}
\label{tab:Y}
\begin{tabular}{lc}
\toprule
Field & $Y$ \\
\midrule
$Q_L=(u_L,d_L)$ & $+1/6$\\
$u_R$ & $+2/3$\\
$d_R$ & $-1/3$\\
$L_L=(\nu_L,e_L)$ & $-1/2$\\
$e_R$ & $-1$\\
$\nu_R$ & $0$\\
\bottomrule
\end{tabular}
\end{table}
Thus the same Cartan directions selected after the holonomy 
breaking contain the conventional Standard-Model hypercharge generator.

\section{Discussion}
The result is summarized by
\begin{align}
N=3
\;&\Longrightarrow\;
\text{closed three-body holonomy cycle} 
\nonumber \\ 
\;&\Longrightarrow\;
3\to2+1 
\;\Longrightarrow\;
\Delta\mathcal M\propto T_8  
\nonumber \\  
\;&\Longrightarrow\;
SU(3)\to SU(2)\times U(1).
\label{eq:chain}
\end{align}
The first arrow is part of the physical content of the construction: 
in one spatial dimension, three particles are the minimal case 
in which the relative configuration space is two-dimensional 
and the sector-matching problem admits a closed cycle around triple coincidence.
The mechanism selects a breaking direction rather than replacing a Higgs sector.  
In a trinification-like embedding, $T_{8L}$ selected by the left-sector holonomy, 
together with $T_{3R}$ and $T_{8R}$ from the right sector, 
spans the hypercharge direction of Eq.~\eqref{eq:Y}.

Three points delimit the claim.  First, $C_3$ is not identified with $SU(3)$; 
the continuous symmetry arises from a degenerate three-dimensional internal space.  
Second, the present model does not determine why the left and right sectors 
choose different holonomies.  Third, Cartan holonomies preserve rank, 
so the reduction to a single $U(1)_Y$ requires an additional condensate, 
boundary condition, or equivalent rank-reducing ingredient.  
These are model-building inputs beyond the holonomy mechanism itself.

\section{Conclusion}
Three is distinguished in this construction before any gauge embedding is imposed: 
it is the minimal particle number that supports the closed holonomy cycle 
in the relative space, and the fixed-parity ordering structure supplies 
a three-dimensional cyclic internal space.  A cyclic three-body holonomy 
then produces a $2+1$ splitting at $\theta_3=0$ whose mass operator is proportional 
to $T_8$.  This directly selects the conventional
\[
SU(3)\to SU(2)\times U(1)
\]
breaking direction.  In a trinification-like embedding, 
the surviving Cartan generators contain
\[
Y=T_{3R}+\frac1{\sqrt3}(T_{8L}+T_{8R}),
\]
and reproduce the Standard-Model hypercharges.  
Additional dynamics are required for rank reduction and vacuum selection.  
The main result is therefore a minimal topological mechanism 
by which a three-body phase can choose the electroweak breaking direction inside 
an $SU(3)$ sector.

\end{document}